\newif\ifbbB\bbBfalse                
\bbBtrue                             

\input harvmac
\overfullrule=0pt

\def\Title#1#2{\rightline{#1}
 \ifx\answ\bigans
  \nopagenumbers\pageno0\vskip1in \baselineskip 15pt plus 1pt minus 1pt
 \else
  \def\listrefs{\footatend\vskip 1in\immediate\closeout\rfile\writestoppt
   \baselineskip=14pt\centerline{{\bf References}}\bigskip{\frenchspacing%
  \parindent=20pt\escapechar=` \input
refs.tmp\vfill\eject}\nonfrenchspacing}
  \pageno1\vskip.8in
 \fi
 \centerline{\titlefont #2}\vskip .5in}

\ifbbB
 \message{If you do not have msbm blackboard bold fonts,}
 \message{change the option at the top of the tex file.}
 \font\blackboard=msbm10 
 \font\blackboards=msbm7 \font\blackboardss=msbm5
 \newfam\black \textfont\black=\blackboard
 \scriptfont\black=\blackboards \scriptscriptfont\black=\blackboardss
 \def\Bbb#1{{\fam\black\relax#1}}
\else
 \def\Bbb{\bf}
\fi

\def\NPB#1#2#3{{\sl Nucl. Phys.} \underbar{B#1} (#2) #3}
\def\PLB#1#2#3{{\sl Phys. Lett.} \underbar{#1B} (#2) #3}

\def\til{\widetilde}
\def\bar{\overline}
\def\bC{{\Bbb C}}
\def\bI{{\Bbb I}}
\def\bJ{{\Bbb J}}
\def\bR{{\Bbb R}}
\def\bZ{{\Bbb Z}}
\def\USp{{U\!Sp}}
\def\t{{}^t\!}
\def\ph#1{\phantom{#1}}
\def\undertext#1{$\underline{\smash{\hbox{#1}}}$}


\nref\lectures{K. Intriligator and N. Seiberg, 
``Lectures on supersymmetric gauge theories and electric-magnetic duality,'' 
hep-th/9509066.}

\nref\Sei{N. Seiberg,
hep-th/9411149, \NPB{435}{1995}{129}.}

\nref\IP{K. Intrilligator and P. Pouliot, hep-th/9505006,
\PLB{353}{1995}{471}.}

\nref\APSei{P.C. Argyres, M.R. Plesser, and N. Seiberg, 
``The moduli space of vacua of N=2 SUSY QCD and duality in N=1 SUSY QCD,'' 
hep-th/9603042.}

\nref\AS{P.C. Argyres and A.D. Shapere, 
hep-th/9509075, \NPB{461}{1996}{463}.} 

\nref\Hanany{A. Hanany, hep-th/9509176, \NPB{466}{1996}{85}.}

\nref\IS{K. Intriligator and N. Seiberg, 
hep-th/9503179, \NPB{444}{1995}{125}.}

\nref\SWi{N. Seiberg and E. Witten, 
hep-th/9407087, \NPB{426}{1994}{19}.} 

\nref\AD{P.C. Argyres and M.R. Douglas, 
hep-th/9505062, \NPB{448}{1995}{93}.} 

\nref\APSW{P.C. Argyres, M.R. Plesser, N. Seiberg, and E. Witten, 
hep-th/9511154, \NPB{461}{1996}{71}.}

\nref\EHIY{T. Eguchi, K. Hori, K. Ito, S.-K. Yang,
``Study of N=2 Superconformal Field Theories in Four Dimensions,''
hep-th/9603002.}

\nref\EH{T. Eguchi and K. Hori,
``N=2 superconformal field theories in 4 dimensions and A-D-E
classification,''
hep-th/9607125.}

\nref\GlobAnom{E. Witten, 
\PLB{117}{1982}{324}.}


\Title{hep-th/9608129, RU-96-61, WIS-96-35, UK-HEP/96-08}
{\vbox{\centerline{$N{=}2$ Moduli Spaces and $N{=}1$ Dualities for} \medskip
\centerline{$SO(n_c)$ and $\USp(2n_c)$ Super--QCD}}}
\bigskip
\centerline{Philip C. Argyres${}^{1,\star}$,
M. Ronen Plesser${}^{2,\dagger}$,
and Alfred D. Shapere${}^{3,\ddagger}$}
\smallskip
\centerline{\it ${}^1$Department of Physics and Astronomy,
Rutgers University, Piscataway NJ 08855 USA}
\centerline{\it ${}^2$Department of Particle Physics,
Weizmann Institute of Science, 76100 Rehovot Israel}
\centerline{\it ${}^3$Department of Physics and Astronomy,
University of Kentucky, Lexington KY 40506}
\centerline{\tt ${}^\star$argyres@physics.rutgers.edu,
${}^\dagger$ftpleser@wicc.weizmann.ac.il,}
\centerline{\tt ${}^\ddagger$shapere@amoeba.pa.uky.edu}
\bigskip\bigskip
\baselineskip 18pt
\noindent
We determine the exact global structure of the moduli space of $N{=}2$
supersymmetric $SO(n)$ and $\USp(2n)$ gauge theories with matter
hypermultiplets in the fundamental representations, using the
non-renormalization theorem for the Higgs branches and the exact
solutions for the Coulomb branches.  By adding an $(N{=}2)$--breaking
mass term for the adjoint chiral field and varying the mass, the
$N{=}2$ theories can be made to flow to either an ``electric'' $N{=}1$
supersymmetric QCD or its $N{=}1$ dual ``magnetic'' version. We thus
obtain a derivation of the $N{=}1$ dualities of \Sei.

\Date{8/96}

\newsec{Introduction and Discussion}

Over the past two years much progress has been made in our understanding
of the vacuum structure of supersymmetric gauge theories (for a review
see \lectures).  One of the
most interesting new phenomena uncovered 
is ``$N{=}1$ duality'' \Sei, where
two different microscopic gauge theories have the same
infrared behavior.  The most striking examples
of $N{=}1$ dual pairs involve one theory which is an asymptotically-free
(AF) non-Abelian gauge theory and
a second, dual, infrared-free theory with a different gauge group.
In such cases the dual theory gives an explicit description of the
low-energy physics at strong coupling. 
The identification of this free gauge group within the context of
the microscopic AF 
theory is difficult.  Indeed, the dual IR-free gauge group is {\it
magnetic} compared to the {\it electric} AF group.

In the case of $SU(n_c)$ $N{=}1$ super-QCD with $n_f$ flavors, the
dual gauge theory $SU(n_f-n_c)$ with $n_f$ flavors has been derived
\APSei\ by flowing down from the $N{=}2$ supersymmetric version of the
theory.  This paper extends that analysis to the $SO(n_c)$ and
$\USp(2n_c)$\foot{Here $\USp(2n)$ denotes the unitary 
symplectic group of rank $n$.}
gauge groups with matter in the defining representations.
In these cases, the dual groups according to \Sei\ and \IP\ are, 
respectively, $SO(n_f-n_c+4)$ (for $n_f {>} n_c {-}2)$ and $\USp(2n_f-2n_c-4)$
(for $n_f {>} n_c {+} 2$). 

The bulk of the paper is concerned with mapping out a global
description of the $N{=}2$ moduli space of these theories.  We do this
by first solving the classical vacuum equations, and then extending
these solutions to the quantum theory using nonrenormalization
arguments as well as the known exact solutions \AS\Hanany\ for the
Coulomb branches.  Along the way we obtain a compact gauge-invariant
description of the $N{=}2$ moduli space, which turns out to be quite a
bit simpler for these gauge groups than for the $SU(n)$ case.

With the $N{=}2$ moduli space in hand, we then break to $N{=}1$
supersymmetry by turning on a bare mass $\mu$ for the $N{=}1$ adjoint chiral
multiplet part of the $N{=}2$ vector multiplet.
If the AF $N{=}2$ theory is characterized by a
strong-coupling scale $\Lambda$, then for $\mu \gg \Lambda$ we flow to
the corresponding microscopic (AF) $N{=}1$ theory.  For $\mu \ll
\Lambda$, on the other hand, we first integrate out the degrees of
freedom with mass of order $\Lambda$, thus effectively recovering the
$N{=}2$ moduli space derived earlier. The small ($N{=}2$)--breaking
mass $\mu$ generically lifts these vacua, except for special points,
among which is the point of maximal, mutually local degeneration.
This point corresponds to an $N{=}2$ vacuum with precisely the dual
(IR-free) gauge group of \Sei\ and \IP.

Flowing to the IR limit in each of the cases $\mu \gg 
\Lambda$ and $\mu \ll \Lambda$, we are led to a
derivation of the $N{=}1$ dualities of \Sei\ and \IP.  Because $N{=}1$
supersymmetry prevents a phase transition between small and large
$\mu$, the IR theories obtained in these two different limits must be
equivalent.  This equivalence goes beyond the the earlier arguments of
\Sei\ and \IP, which essentially only found an equivalence between the
chiral rings of the two theories.  However, in at least one respect,
our argument is incomplete: we obtain no information on the extra
gauge singlet fields appearing in the dual theories \Sei\ and \IP.

One striking feature of the corresponding analysis in the $SU(n)$ case
\APSei\ was that the IR-free $N{=}1$ dual gauge theory can be
continuously deformed through the larger $N{=}2$ moduli space to a 
theory whose gauge group is a subgroup
of the microscopic (AF) gauge group.  The existence of a continuous
interpolation between electric and magnetic degrees of freedom is
allowed because there is no phase transition separating Higgs and
confining phases (condensation of electric and magnetic charges,
respectively) in $SU(n)$ with fundamental scalars.  In $SO(n)$ with
matter in the vector ($\bf n$) representation, however, there is such
a distinction, since a Wilson loop in a spinor representation cannot
be screened by vector charges.  Thus, in the $SO(n)$ case one expects
that any such interpolation must pass through a phase transition.

This is indeed what we find: the branches of the $SO(n)$ $N{=}2$
moduli space that connect to the IR-free non-Abelian vacuum in
question all have unbroken $U(1)$ gauge factors.  So as we deform
$N{=}1$ electric Higgs or confining (magnetic Higgs) vacua to the
corresponding $N{=}2$ vacua, they gain $U(1)$ factors and so the
asymptotic behavior of spinorial Wilson loops in these vacua changes
abruptly to Coulomb-like behavior as $N{=}2$ symmetry is restored.
This is because the charges in such a Wilson loop are charged under
all the $U(1)$ factors in the Cartan subalgebra of the microscopic
$SO(n)$ group---the spinor weights of $SO(n)$ are
$(\pm\half,\ldots,\pm\half)$.

However, the observation that the electric and magnetic Higgs phases
are distinct does not invalidate our duality argument.  Because there is
no electric/magnetic phase transition in the Coulomb phase, we
{\it can} interpolate continuously between the dual IR theories,
provided that during the interpolation we remain in the Coulomb phase
at all times.

\newsec{Classical $N{=}2$ $SO(n_c)$ Moduli Space}

$N{=}2$ supersymmetric $SO(n_c)$ Yang-Mills theory is described in terms 
of $N{=}1$ superfields by a field strength chiral multiplet $W^\alpha_{ab}$ 
and a scalar chiral multiplet $\Phi_{ab}$, both in the adjoint representation 
of the gauge 
group, which together form an $N{=}2$ vector multiplet.  Here $a,b=1,\ldots,
n_c$ are color indices.  The Lagrangian is
                \eqn\Ni{
{\cal L}_{\rm YM} = {\rm tr\ Im}\Biggl[\tau\!\int\!\! d^2\theta\,
d^2\bar\theta\,\, \bar\Phi e^V \Phi +
\tau\!\int\!\! d^2\theta\, {\textstyle{1\over2}}{\rm tr}W^2 \Biggr],
                }
where $\tau$ is the gauge coupling and theta angle $\tau = (\theta/\pi) + i 
(8\pi/g^2)$.

Matter in the $\bf n_c$ representation of the gauge group is made up of the
$N{=}1$ chiral ``quark'' multiplets $Q^i_a$, $i=1,\ldots,2n_f$, pairs of which
$(Q^i_a,Q^{i+n_f}_a)$ together make up $N{=}2$ hypermultiplets.  Matter couples
to the Yang-Mills fields {\it via} the usual kinetic terms and a cubic
superpotential:
		\eqn\Nii{
{\cal L}_{\rm matter} = \int\!\! d^2\theta\, d^2\bar\theta\,\, 
\bar Q^i_a (e^V)_{ab} Q^i_b +
\int\!\! d^2\theta\, \sqrt2 Q_a^i \Phi_{ab} Q_b^j \bJ^{ij} + c.c. ,
		}
where $\bJ$ is the symplectic metric $\bJ \equiv {\ 0\ 1\choose-1\ 0}
\otimes \bI_{n_f \times n_f}$, and $\bI$ is the identity matrix.

The theory has a global $\USp(2n_f)$ flavor symmetry (when there are no
bare quark masses) as well as a $U(1)_R \times SU(2)_R$ chiral
R-symmetry.  $\USp(2n_f)$ is the group of $2n_f \times 2n_f$ complex
matrices $M$ satisfying $M\cdot \bI\cdot \t\bar M = \bI$ and $M\cdot
\bJ\cdot \t M = \bJ$, and thus preserving both the hermitian $Q_a \cdot \bar
Q_b$ and symplectic $Q_a \cdot_J Q_b$ inner products of complex
$2n_f$-component vectors.  Mass terms and instanton corrections break
$U(1)_R$.  When $n_f < n_c {-} 2$, the theory is asymptotically free
and generates a strong-coupling scale $\Lambda$, the instanton factor
is proportional to $\Lambda^{2n_c-2n_f-4}$, and the $U(1)_R$ symmetry
is anomalous, being broken by instantons down to a discrete
$\bZ_{2n_c-2n_f-4}$ symmetry.  For $n_f=n_c {-} 2$ the theory is
scale-invariant and the $U(1)_R$ is not anomalous.  In this case no
strong coupling scale is generated, and the theory is described in
terms of its bare coupling $\tau$.

The classical vacua are the zeroes of the scalar potential,
found by setting the $D$- and $F$-terms to zero. The $D$-term 
equations are
		\eqn\Dvaceqs{\eqalign{
0 &= [\Phi,\bar\Phi] ,\cr
0 &= {\rm Im} ( Q_a \cdot \bar Q_b ),
		}}
and the $F$-term equations are
                \eqn\Fvaceqs{\eqalign{
0 &= Q_a \cdot_J Q_b ,\cr
0 &= \Phi^{\ph i}_{ab} Q^i_b .
                }}
These vacuum equations imply that the fields 
$\Phi$ and $Q$ may get vevs, which we
denote by the same symbols.  The solutions to the vacuum equations form
various ``branches'' corresponding to different phases of the
theory.  The Coulomb branch is defined as the set of solutions with
$Q=0$, Higgs branches are those with $\Phi=0$, and mixed branches are
those with both $\Phi$ and $Q$ nonvanishing.

\undertext{\it Coulomb Branch:}  The Coulomb branch satisfies
$[\Phi,\bar\Phi]=0$ with $Q=0$, implying that $\Phi$ can be
skew-diagonalized by a color rotation to a complex matrix
		\eqn\Ci{
\Phi=\pmatrix{
0      &\phi_1&      &       &      \cr
-\phi_1&0     &      &       &      \cr
       &      &\ddots&       &      \cr
       &      &      &0      &\phi_{[n_c/2]}\cr
       &      &      &-\phi_{[n_c/2]}&0     \cr}.
		}
This vev generically breaks $SO(n_c)\rightarrow
U(1)^{[n_c/2]}$, motivating the name for this branch.

For $n_c$ even, gauge transformations in the Weyl group of $SO(n_c)$
are generated by permutations, and by simultaneous sign changes of pairs
of the $\phi_a$.  So the symmetric polynomials $S_\ell$ of the
$\phi_a^2$, generated by $\sum_\ell S_\ell x^{[n_c/2]-\ell} = \prod_a
(x {-} \phi_a^2)$, are ``glue'' gauge invariants.  In addition to the
$S_\ell$, there is one ``extra'' Weyl invariant $T = 
\phi_1 \phi_2 \cdots \phi_{n_c/2} = \pm\sqrt{S_{n_c/2}}$.  

For $n_c$ odd, there is an extra
row and column of zeroes in \Ci.  The Weyl group is generated by
permutations and individual sign flips of the $\phi_a$, so the
symmetric polynomials $S_\ell$ in the $\phi_a^2$ form a complete basis
of glue invariants on the Coulomb branch.

This classical moduli space has orbifold singularities along
submanifolds where some of the $\phi_a$'s are equal or vanish.  In this
case some of the non-Abelian gauge symmetry is restored.  If $k$
$\phi_a$'s are equal and non-zero, there is an enhanced $SU(k)$ gauge
symmetry.  If they are also zero, then there is an enhanced $SO(2k)$ or
$SO(2k{+}1)$, depending on whether $n_c$ is even or odd, respectively.
In this case, the glue invariants $S_\ell =0$ for $\ell > [n_c/2] {-}
k$. 

\undertext{\it Higgs Branches:} The Higgs branch is the space of
solutions to the second equation in \Dvaceqs\ and the first in
\Fvaceqs\ since $\Phi=0$.  Describe the squark fields as complex
matrices with $n_c$ rows and $2n_f$ columns.  Any solution of the Higgs
branch equations can be put in the following form by a combination of
flavor and gauge rotations:
		\eqn\Hii{
Q = \pmatrix{
q_1      &&& \ph{q_1}    &&\cr 
&\ddots   && &\ph\ddots   &\cr
&&q_{n_f}  & &&\ph{q_{n_f}}\cr
\ph\vdots&&& &&            \cr} 
,\qquad q_a\in\bR^+ .
		}
In \Hii\ we assumed $n_c > n_f$; if $n_c < n_f$, then there will be
$n_c$ entries on the diagonal.  Call such a solution with $r$ of the
$q_i$ non-zero the $r$-Higgs branch.  Thus, on the $r$-Higgs branch an
$SO(n_c{-}r)$ gauge symmetry is unbroken, with $n_f{-}r$ massless
hypermultiplets transforming in its vector representation.  So, by the
Higgs mechanism, of the $n_f n_c - (n_f{-}r) (n_c{-}r)$ neutral
hypermultiplets, ${1\over2} n_c(n_c{-}1) - {1\over2} (n_c{-}r)
(n_c{-}r{-}1)$ are given a mass, leaving ${\cal H} = rn_f -
{1\over2}r(r{-}1)$ massless neutral hypermultiplets---the quaterionic
dimension of the $r$-Higgs branch.  (As we will see later, this
counting is really only accurate for $n_c{-}r$ even.)

In order to identify the unbroken global symmetries on the $r$-Higgs
branches, it is useful to define a basis of gauge-invariant quantities
made from the squark vevs, the meson and baryon fields
		\eqn\mesbary{\eqalign{
M^{ij} &= Q_a^i Q_a^j ,\cr
B^{[i_1\ldots i_{n_c}]} &= Q^{i_1}_{a_1}\cdots Q^{i_{n_c}}_{a_{n_c}}
\epsilon_{a_1\ldots a_{n_c}}\ .
		}}
The baryon field is defined for $2n_f \ge n_c$.  {}From our solution
for the $r$-Higgs branch squark vevs we see $B \neq 0$ only when $r =
n_c \le n_f$.  The meson field is diagonal with $r$ $q_i^2$'s along the
diagonal.  It therefore leaves a $\USp(2n_f{-}2r)$ global flavor
symmetry unbroken.  (A non-vanishing baryon field does not break this
symmetry.) Thus the number of real goldstone modes is ${\cal G} =
n_f(2n_f {+}1) - (n_f{-}r) (2n_f{-}2r{+}1)$, and the number of real
parameters describing the Higgs branch is ${\cal P} = r$.  It is a
check on our counting that ${\cal G} {+} {\cal P} = 4{\cal H}$.

\undertext{\it Mixed Branches:} Using the antisymmetry of $\Phi$, it
follows from the second $F$-term equation in \Fvaceqs\ that on an
$r$-Higgs branch $\Phi$ must be zero except in a lower right-hand
$(n_c{-}r) \times (n_c{-}r)$ block, which can be skew diagonalized by
the unbroken $SO(n_c{-}r)$ rotations.  Call the mixed branch which
emanates from an $r$-Higgs branch simply the $r$-branch.  When
$n_c{-}r$ is odd, there will be a row and column of zeros in $\Phi$,
with a corresponding row of zeros in $Q$.  Thus, such an $r$-branch is
really just a submanifold of the $r{+}1$-branch, and not a separate
branch.  {}From now on we denote by ``$r$-branches'' only those with
$n_c{-}r$ even.\foot{There is an exception:  if $n_c{-}n_f$ is odd, then
for the maximal value of $r$, $r{=}n_f$, there is no $n_f{+}1$--branch
for the $n_f$--branch to be a submanifold of.}

With non-zero vevs for $\Phi$ and $Q$ we can define, in addition to the
glue invariants $S_\ell$ and meson $M$, a set of ``baryonic'' invariants
		\eqn\baryinvs{
B_\ell^{[i_1 \ldots i_{n_c-2\ell}]} = \Phi^{\ph{i_1}}_{a_1 a_2} \cdots
\Phi^{\ph{i_1}}_{a_{2\ell-1}a_{2\ell}} Q^{i_1}_{a_{2\ell+1}}\cdots
Q^{i_{n_c-2\ell}}_{a_{n_c}} \epsilon^{\ph{i_1}}_{a_1 \cdots a_{n_c}}.
		}
Note that $\ell=0$ corresponds to the usual baryon, while $\ell =
n_c/2$ (for $n_c$ even) gives the ``extra'' Coulomb-branch invariant
$T$.  {}From the block form of the $Q$ and $\Phi$ vevs found above on
the $r$-branch, it follows that all the baryon invariants vanish except
for $B_{(n_c-r)/2}$.  Also, the glue invariants $S_\ell=0$ for $\ell >
(n_c{-}r)/2$.

\undertext{\it Gauge-Invariant Description of Moduli Space:} We have
found above only representative solutions for the $Q$ and $\Phi$ vevs,
since global symmetry transformations on these solutions will relate
them to distinct points in the moduli space.  To have a global
description, it is useful to describe the various branches in terms of
constraints on the gauge-invariant meson, glue, and baryon order
parameters.  In particular, setting the $D$-terms to zero and
identifying orbits of the gauge group is equivalent to dividing out
the space of $Q$ and $\Phi$ vevs by the action of the 
complexified gauge group. The latter operation may be 
achieved by expressing the vevs in terms of holomorphic gauge-invariant
coordinates, which, however, are not independent as functions of the
$Q$ and $\Phi$ vevs, but satisfy a set of polynomial relations.  Below
we find a set of generators for the constraints following from these
relations and the $F$-term equations.

Since the product of two color epsilon-tensors is the antisymmetrized
sum of Kronecker deltas, it follows that
		\eqn\mbconstri{
B_k\, B_\ell = \delta_{k\ell}\,  S_\ell\,\, {*}(M^{n_c-2\ell}),
\qquad 0 \le \ell \le [n_c/2],
		}
where the ``$*$'' denotes the antisymmetrization of the product of
$M$'s on half their flavor indices, and flavor indices are uncontracted.  
Note that the $F$-term equation
$\Phi \cdot Q =0$ has been used in deriving \mbconstri.  Also, since
any expression antisymmetrized on $n_c{+}1$ color indices must vanish,
it follows that any product of $M$'s and $B$'s antisymmetrized on
$n_c{+}1$ flavor indices must vanish.  \mbconstri\ can be used to
eliminate all the $B$'s from such constraints, leading to one other
independent constraint $*(M^{n_c{+}1}) = 0$.  Another set of
independent constraints follows from contracting the color identity $0 =
\delta^{a_1}_{[b_1} \cdots \delta^{a_{n_c+1}}_{b_{n_c+1}]}$ with
$2\ell$ $\Phi$'s and $(2n_c {+} 2 {-} 4\ell)$ $Q$'s.  Using $\Phi\cdot
Q=0$ one then finds
		\eqn\mbconstrii{
0 = S_\ell\,\, *(M^{n_c-2\ell+1}),\qquad 0\le \ell \le [n_c/2].
		}
Note that when $\ell=0$ ($S_0 \equiv 1$) this is equivalent to
$*(M^{n_c{+}1}) = 0$.

The $F$-term equation $Q \cdot_J Q = 0$ implies the further constraint
		\eqn\mbconstriii{
M \cdot_J M = 0.
		}
The other constraints following from the $F$-term equation, $B_\ell
\cdot_J M = B_\ell \cdot_J B_k = 0$, are not independent of
\mbconstri\ and \mbconstriii.  Thus \mbconstri, \mbconstrii, and
\mbconstriii\ form a complete set of constraints describing the
classical moduli space.

We can solve these constraints and recover the properties of the
$r$-branches found above.  Eq.\ \mbconstriii\ implies $M$ can be
diagonalized to $r$ positive real entries with $r \le n_f$.\foot{To see
this, note that \mbconstriii\ implies that the image of $M$ (viewed as a 
linear transformation) is
symplectically orthogonal; in particular, $r \equiv {\rm dim}({\rm
im}\,M) \le n_f$.  So we can choose an orthonormal basis of im$M$ and
extend it to an ortho-symplectic basis of the full space.  Thus there
is a $\USp(2n_f)$ similarity transformation expressing $M$ in this
basis where its last $2n_f{-}r$ columns, and hence rows by its
symmetry, vanish.  It follows that $M$ is zero except for a complex symmetric
upper left-hand $r\times r$ block, which can be diagonalized to
non-negative real entries by a $U(r) \subset \USp(2n_f)$ similarity
transformation.}  Furthermore, from \mbconstrii\ with $\ell{=}0$ we
learn that actually $r \le {\rm min}\{n_f,n_c\}$, reproducing the form
of the meson fields on the $r$-branches.  By \mbconstri, if any one
baryon invariant does not vanish, say $B_\ell \neq 0$, then all the
other $B_k=0$ for $k\neq\ell$.  Since on an $r$-branch rank$(M)=r$, we
have that $*(M^{n_c-2k})=0$ for $k < (n_c{-}r)/2$ and are non-vanishing
otherwise.  Then \mbconstrii\ implies that $S_k =0$ for $k>
(n_c{-}r)/2$, and from \mbconstri\ we learn that the one non-vanishing
$B_\ell$ must have $\ell = (n_c{-}r)/2$ (for $n_c{-}r$ even).

\undertext{\it Summary}:  We have found that the $SO(n_c)$ theory with
$n_f$ vector flavors has a moduli space made up of $r$-branches with $0
\le r \le {\rm min}\{n_f, n_c\}$ with $n_c{-}r$ even.  The $r$-branch
has hypermultiplet dimension ${\cal H}= rn_f - {1\over2}r(r{-}1)$ and
vector multiplet dimension ${\cal V} = {1\over2} (n_c {-} r)$.  Thus,
the $(r{=}0)$-branch is the Coulomb branch, the $(r{=}1)$-branch 
includes the Coulomb
branch as a submanifold, while for $r{=}n_c$ we obtain a pure Higgs
branch.  The ``root'' of an $r$-branch is its submanifold of
intersection with the Coulomb branch.  Thus, the $r$-branch root has
quaternionic dimension $(n_c{-}r)/2$ and has an $SO(r)\times U(1)^{(n_c-r)/2}$
unbroken gauge group classically.

\newsec{Quantum $N{=}2$ $SO(n_c)$ Moduli Space}

A non-renormalization theorem \APSei\ implies that quantum mechanically the
$r$-branches retain their Coulomb $\times$ Higgs product structure, the
Higgs factors are not renormalized and do not depend on the quark
masses, and the Coulomb factors are given by submanifolds of the
quantum Coulomb branch.  

We have seen that classically there exist $r$-branches for $0 \le r
\le n_f$ with $n_c {-} r$ even which meet the Coulomb branch along
submanifolds with gauge group $SO(r)\times U(1)^{(n_c-r)/2}$ with
$n_f$ vector flavors.  Since $SO(n_c)$ gauge theories are only
asymptotically free when $n_f \le n_c {-} 2$, the $SO(r)$ factors at
the roots of the $r$-branches are all IR-free, and will remain
unbroken quantum-mechanically.

Submanifolds of the quantum Coulomb branch with unbroken $SO(r)$ gauge
factors are easy to identify explicitly using the exact solution for
the Coulomb branch found in \AS.  The generic vacuum on the Coulomb
branch is a $U(1)^{[n_c/2]}$ pure Abelian gauge theory characterized by
an effective coupling $\tau_{ij}$ between the $i$th and $j$th $U(1)$
factors, which, due to the ambiguity of electric-magnetic duality
rotations in the $U(1)$ factors, forms a section of an
$Sp(2[n_c/2],\bZ)$ bundle over the Coulomb branch \SWi.  An explicit
description of the Coulomb branch is given by associating to each point
of the Coulomb branch a genus $[n_c/2]$ Riemann surface whose complex
structure is the low energy coupling $\tau_{ij}$.  Globally the quantum
Coulomb branch can still be characterized by $[n_c/2]$ complex numbers
$\phi_a$ (up to permutations and sign flips) just as in the classical
analysis of the last section.  The family of Riemann surfaces
describing the effective action on the Coulomb branch with $n_f$
massless flavors is then \AS
		\eqn\Escalecurv{
y^2 = x\prod_{a=1}^{[n_c/2]}(x-\phi^2_a)^2 - 4\Lambda^{2(n_c-2-n_f)}
x^{n_f+2+\epsilon},
		}
where $\epsilon = 1$ if $n_c$ is even, and $\epsilon = 0$ if $n_c$ is
odd.  This form of the solution is valid for all AF values of $n_f$.
In the finite case, when $n_f = n_c {-} 2$, $\Lambda^0$ should be
replaced by a known function of the bare coupling.  In the IR-free
case, when $n_f > n_c {-} 2$, \Escalecurv\ is valid in a sufficiently
small neighborhood of $x=\phi_a=0$ \APSei.  In particular, in the
IR-free case, the form of the curve at the origin of moduli space
(where the $SO(n_c)$ gauge symmetry is restored) is simply
		\eqn\IRfreecrv{
y^2 = x^{n_c+\epsilon}(1 - 4\Lambda^{2(n_c-2-n_f)} x^{n_f-n_c+2}),
\qquad n_f > n_c{-}2.
                }

When two or more of the branch points of \Escalecurv\ collide as we
vary the moduli, the Riemann surface degenerates, giving a singularity
in the effective action corresponding to additional $N{=}2$ multiplets
becoming massless.  When $n_s$ independent pairs of branch points
collide there will be generically $n_s$ hypermultiplet states becoming
massless (with $U(1)$ charges proportional to the homology classes of
the vanishing cycles on the Riemann surface).  More complicated
singularities will generally lead to different physics.  

In particular, from \IRfreecrv\ we see that when $r$ branch points
coincide, one may expect an unbroken $SO(r)$ or $SO(r{-}1)$ gauge
symmetry.  Such singularities are easy to find in the AF curves
\Escalecurv\ with $n_f \le n_c {-} 2$:  just set some of the
$\phi_a=0$.  Thus, on the submanifold with all but $(n_c{-}r)/2$ of
the $\phi_a=0$ (where $n_c{-}r$ is even), the curve becomes
		\eqn\rootcrv{
y^2 = x^{r+\epsilon}\left[\prod_{a=1}^{(n_c-r)/2}(x-\phi_a^2)^2 -
4\Lambda^{2(n_c-2-n_f)}x^{n_f-r+2}\right],
		}
suggesting vacua with an unbroken $SO(r)\times U(1)^{(n_c-r)/2}$ gauge
group.  This interpretation is confirmed by the fact that these
singular submanifolds reach far out on the Coulomb branch ($\phi_a \gg
\Lambda$) where they have the semi-classical interpretation as the
submanifolds of the Coulomb branch where an IR-free $SO(r)\times
U(1)^{(n_c-r)/2}$ group is left unbroken.

We have thus located the roots of the $r$-branches in the full quantum
theory, and found that the structure of the quantum moduli space is
qualitatively much the same as its classical structure.  It is easy to
check that the IR--free vacua at the $r$-branch roots indeed have mixed
Higgs-Coulomb branches emanating from them which are precisely the same
as the $r$-branches determined classically in the last section.

Since the theories at the $r$-branch roots are IR--free, their gauge
symmetry will survive quantum-mechanically.  Quantum effects could,
however, change this effective theory by bringing down additional light
degrees of freedom.  In particular, there may be points on the
$r$-branch root submanifolds where (monopole) singlets charged under
the $U(1)$ factors become massless.  Such points are located where the
factor in square brackets in \rootcrv\ becomes singular due to pairs of
its zeros coinciding.  The maximal such singularity occurs when that
factor is a perfect square, corresponding to $(n_c{-}r)/2$
hypermultiplets becoming simultaneously massless.  We will see in the
next section that these vacua are especially interesting since they
remain vacua upon breaking to $N{=}1$ supersymmetry.

Expanding out the terms in square brackets in \rootcrv, the condition
that they form a perfect square is
		\eqn\perfsqr{\eqalign{
\Bigl[x^{(n_c-r)/2} + s_1 x^{(n_c-r)/2 -1} + \ldots & 
+ s_{(n_c-r)/2}\Bigr]^2 - 4\Lambda^{2(n_c-2-n_f)} x^{n_f-r+2}\cr
&= \left[x^{(n_c-r)/2} + \til s_1 x^{(n_c-r)/2 -1} + \ldots + 
\til s_{(n_c-r)/2} \right]^2 \cr
		}}
for some $s_\ell$ and $\til s_\ell$.  Moving the first term on the left
to the right and factorizing, it is then easy to show that if
$r > 2n_f {-} n_c {+} 4$ or $ n_f \ge n_c{-}2$
there is no solution, and if $r \le 2n_f
{-} n_c {+} 4$ the only solution is $s_{n_c-2-n_f} =
\Lambda^{2(n_c-2-n_f)}$ with all the other $s_\ell =0$.  Plugging this
solution into \rootcrv\ gives the curve
		\eqn\specialcrv{
y^2 = x^{2n_f - n_c + 4 + \epsilon} \left( x^{n_c -2 -n_f} -
\Lambda^{2(n_c-2-n_f)} \right)^2,
		}
in which $r$ has dropped out.  Thus, we have located the unique point
on the $SO(n_c)$ Coulomb branch with $SO(r)\times U(1)^{(n_c-r)/2}$
unbroken IR gauge group and the maximal number $(n_c {-} r)/2$ of singlets
charged under the $U(1)$'s.  By comparison with \rootcrv\ we see that
\specialcrv\ corresponds to $r= 2n_f {-} n_c {+}4$ and
		\eqn\specialpnt{
\phi_a^2 = \Lambda^2 (0, \ldots, 0 , \omega, \omega^2, \ldots,
\omega^{n_c-2-n_f})
		}
where $\omega = {\rm exp}\{2\pi i/(n_c{-} 2{-}n_f)\}$.

A simple contour argument shows that we can pick a basis in which the 
singlets have a diagonal charge matrix, with each singlet having charge 
$1$ under only one of the $U(1)$'s.  The squarks are neutral under the
$U(1)$'s since they are in a real flavor representation.
These charges can be summarized as follows:
                \eqn\macro{\matrix{
&SO(2n_f{-}n_c{+}4)&\times&U(1)_1&\times&\cdots&\times&U(1)_{n_c-2-n_f}\cr
2n_f\times Q  &\bf 2n_f{-}n_c{+}4 && 0      && \cdots && 0      \cr
e_1           &\bf 1              && 1      && \cdots && 0      \cr
\vdots        &\vdots             && \vdots && \ddots && \vdots \cr
e_{n_c-2-n_f} &\bf 1              && 0      && \cdots && 1      \cr
                }}

\newsec{Breaking $N{=}2$ $SO(n_c)$ to $N{=}1$}

In this section we break to $N{=}1$ supersymmetry by turning on bare
masses for the adjoint superfield $\Phi$. 
Since $\Phi$ is part of the $N{=}2$ vector
multiplet $(\Phi,W_\alpha)$, giving it a mass explicitly breaks
$N{=}2$ supersymmetry.  In the microscopic theory, this corresponds to
an $N{=}1$ theory with a superpotential
		\eqn\Nisupotl{
{\cal W} = \sqrt2\, Q_a \cdot_J Q_b \Phi_{ab} - {\mu \over 2} \Phi_{ab}
\Phi_{ab}.
	       }
For $\mu\gg\Lambda$ we can integrate $\Phi$ out in a
weak-coupling approximation, obtaining an effective superpotential
that vanishes as $\mu\rightarrow\infty$. We are thus left with $N{=}1$
$SO(n_c)$ super--QCD with $2n_f$ flavors\foot{In $N{=}1$ theories
we count flavors by the number of squark chiral multiplets.
Thus, by this counting the $N{=}2$ theory with $n_f$ hypermultiplets
has $2n_f$ $N{=}1$ flavors.} and no superpotential at scales above the
strong-coupling scale $\Lambda_1$ of the $N{=}1$ theory.  If the strong
coupling scale of the $N{=}2$ theory is $\Lambda$, then by a one-loop
matching, the $N{=}1$ scale is $\Lambda_1^{3(n_c-2) - 2n_f} =
\mu^{n_c-2} \Lambda^{2(n_c-2) - 2n_f}$.  The appropriate scaling
limit sends $\mu \rightarrow \infty$ and $\Lambda \rightarrow 0$
keeping $\Lambda_1$ fixed, so the model is described by the $N{=}1$
theory on scales between $\mu$ and $\Lambda_1$, below which the
strongly-coupled dynamics of the $N{=}1$ theory takes over.

We can also study the breaking to $N{=}1$ by beginning with
$\mu\ll\Lambda$.  In this case we should study the effects of $\mu$
on the low-energy $N{=}2$ theory obtained in the previous
sections.  $N{=}1$ supersymmetry prevents a phase transition as we vary
$\mu$, hence we should obtain the same result as that obtained
for $\mu\gg\Lambda$.

We will now explain why generic vacua of the $N{=}2$ theory are lifted
by nonzero $\mu$, and also why the special point we found on the
$r=2n_f{-}n_c{+}4$ $r$-branch is not.
We thus study the effects of the
breaking to $N{=}1$ in the effective theories at the roots of the
$r$-branches, which we saw have unbroken gauge groups of the form
$SO(r) \times U(1)^{(n_c-r)/2}$.  Let $\phi$ denote the adjoint scalar
in the $SO(r)$ vector multiplets, and $\psi_k$ the adjoint scalars for
each of the $U(1)$ vector multiplets.  Then the microscopic mass term
$(\mu/2){\rm tr}\Phi^2$ becomes $\mu(\Lambda\sum_i x_i \psi_i +
{1\over2}{\rm tr}\phi^2 + \ldots)$, where the dots denote higher-order
terms, and $x_i$ are dimensionless numbers.  (From the $\Phi$ vev
\specialpnt\ at the special point, we see that all $x_i \sim 1$.)

Note that at any point on an $r$-branch root for which there are fewer
than $(n_c{-}r)/2$ massless singlets, $e_k$, charged under the
$U(1)$'s, then the $N{=}2$ vacuum is lifted.  This can be seen as
follows.  If there are $n_s$ singlets with $n_s < (n_c{-}r)/2$, a basis
of the $U(1)$'s can be chosen to diagonalize the charges of the
singlets, and the superpotential becomes
		\eqn\Ninbbrsp{
{\cal W} = \sqrt2\, {\rm tr}(Q \cdot_J Q \phi) + \sqrt2\sum_{k=1}^{n_s}
\psi_k e_k \til e_k + \mu\Bigl( \Lambda \sum_{i=0}^{(n_c-r)/2} x_i \psi_i
+ {1\over 2} {\rm tr} \phi^2 \Bigr).
		}
The $F$-term equations following from taking derivatives with respect
to the $\psi_i$ then have no solution.  

Therefore only the special
vacuum \macro\ will lead to an $N{=}1$ vacuum.\foot{It may be that
certain other $N{=}2$ vacua corresponding to non-trivial fixed points
\refs{\AD,\APSW,\EHIY,\EH} can also remain $N{=}1$ vacua.} 
In this case the $e_k$
all get vevs, Higgsing all the $U(1)$ factors.  Thus, when $\mu \neq 0$
the $e_k$ and $\psi_i$ fields are massive and can be integrated out,
leaving the effective superpotential
		\eqn\Ninbbreff{
{\cal W}^\prime = \sqrt2\,{\rm tr}(Q\cdot_J Q\phi) + {\mu\over2}{\rm
tr}\phi^2,
		}
for an $N{=}1$ $SO(2n_f{-}n_c{+}4)$ super--QCD with $2n_f$ flavors.
This is precisely the dual gauge group for an even number of flavors of
\Sei.

We should re-emphasize that the arguments given here show that the
microscopic (AF) theory is IR-equivalent to another theory with the
derived dual gauge group, whereas the earlier arguments of \Sei\
essentially only showed this to be true for the chiral rings of the
two theories.  However, we obtain no information on the extra gauge
singlet fields appearing in the dual theory found in \Sei.  Also, much
of the rich structure \refs{\Sei,\IS} of the $N{=}1$ $SO(n)$ moduli
spaces concerning the interplay of their Higgs and confining branches
is missed in our analysis.  Presumably a similar analysis including
bare quark masses would enable us to recover much of this information.

\newsec{$N{=}2$ Moduli Space and $N{=}1$ Duality for $\USp(2n_c)$}

In $N{=}2$ supersymmetric $\USp(2n_c)$ QCD, matter in the $\bf 2n_c$
representation of the gauge group is made up of the $N{=}1$ chiral
``quark'' multiplets $Q^i_a$, $i=1,\ldots,2n_f$, pairs of which
$(Q^{2i-1}_a,Q^{2i}_a)$ together make up $N{=}2$ hypermultiplets, and
which couple to the Yang-Mills fields as
		\eqn\SpNii{
{\cal L}_{\rm matter} = \int\!\! d^2\theta\, d^2\bar\theta\,\, \bar
Q^i_a (e^V)^{ab} Q^i_b + \int\!\! d^2\theta\, \sqrt2 Q_a^i \Phi^{ab}
Q_b^i + c.c. ,
		}
where the symplectic metric $\bJ_{ab} \equiv {\ 0\ 1\choose-1\ 0}
\otimes \bI_{n_c \times n_c}$ is used to raise and lower $\USp(2n_c)$
color indices.  Classically (and with no masses) the theory has a
global $O(2n_f)\simeq SO(2n_f) \times \bZ_2 $ flavor symmetry as well as a
$U(1)_R \times SU(2)_R$ chiral R-symmetry.  Mass terms and instanton
corrections break $U(1)_R$ and the $\bZ_2$ of the flavor symmetry.
When $n_f < 2n_c {+} 2$, the theory is asymptotically free and
generates a strong-coupling scale $\Lambda$, the instanton factor is
proportional to $\Lambda^{2n_c+2-n_f}$, and the $U(1)_R$ symmetry is
anomalous, being broken by instantons down to a discrete
$\bZ_{2n_c+2-n_f}$ symmetry.

The classical vacua are the solutions to the $D$-term equations,
		\eqn\SpDvaceqs{\eqalign{
0 &= \Phi_{ab}\bar\Phi^b_c + \Phi_{cb}\bar\Phi^b_a,\cr
0 &= Q_a \cdot \bar Q_b + Q_b \cdot \bar Q_a,
		}}
and the $F$-term equations,
                \eqn\SpFvaceqs{\eqalign{
0 &= Q_a \cdot Q_b ,\cr
0 &= Q^i_a \Phi^a_b.
                }}
The Coulomb branch satisfies the first $D$-term equation,
implying that $\Phi$ can be diagonalized by a color rotation to
		\eqn\SpCi{
\Phi=\pmatrix{1&0\cr0&-1\cr}\otimes\pmatrix{
\phi_1&      &          \cr
      &\ddots&          \cr
      &      &\phi_{n_c}\cr}, \qquad \phi_a\in \bC,
		}
breaking $\USp(2n_c)\rightarrow U(1)^{n_c}$, except when $k{>}1$ of the
$\phi_a$'s are equal or vanish, in which case an $SU(k)$ or $\USp(2k)$
gauge symmetry is restored, respectively.  The Weyl group of
$\USp(2n_c)$ is generated by permutations and by sign flips of the
$\phi_a$, so the symmetric polynomials $S_\ell$, $\ell = 1, \ldots,
n_c$ in $\phi_a^2$ are ``glue'' gauge invariants.  Along submanifolds
of enhanced $\USp(2k)$ symmetry, $S_\ell =0$ for $\ell > n_c {-} k$.

The Higgs branches comprise the space of solutions to the second
$D$-term equation and the first $F$-term equation since $\Phi=0$.
Describing the squark fields as complex matrices with $2n_c$ rows and
$2n_f$ columns, any solution of the Higgs branch equations can be put
in the following form by a combination of flavor and gauge rotations:
		\eqn\SpHii{
Q = \pmatrix{1&i&&\cr&&1&i\cr}\otimes\pmatrix{
q_1       &&&\cr 
&\ddots    &&\cr
&&q_r       &\cr
&&&\ph\ddots \cr} ,\qquad q_i\in\bR^+ ,
		}
where $r \le {\rm min}\{ n_c, n_f/2 \}$.  On this $r$-Higgs branch a
$\USp(2n_c{-}2r)$ gauge symmetry is unbroken, with $n_f{-}2r$ massless
hypermultiplets transforming in its fundamental representation.  By the
Higgs mechanism, of the $2 n_c n_f - 2(n_c{-}r) (n_f{-}2r)$ neutral
hypermultiplets, $n_c(2n_c{+}1) - (n_c{-}r) (2n_c{-}2r{+}1)$ are given
a mass, leaving ${\cal H} = 2rn_f - r(2r{+}1)$ massless neutral
hypermultiplets.

Gauge-invariant quantities made from the squark vevs can all be made from
the meson field $M^{ij} = Q_a^i \bJ^{ab} Q_b^j$, since the antisymmetric 
tensor on $2n_c$ color indices is just the exterior product of $n_c$ 
symplectic metrics.  By \SpHii, the meson field is 
		\eqn\Spmeson{
M = \pmatrix{
 &  &-1&-i\cr
 &  &-i& 1\cr
1& i&  &  \cr
i&-1&  &  \cr}\otimes\pmatrix{
q_1^2     &&&\cr 
&\ddots    &&\cr
&&q_r^2     &\cr
&&&\ph\ddots \cr} ,
		}
therefore leaving unbroken an $SU(2)^r \times SO(2n_f{-}4r)$ global
flavor symmetry.\foot{The three $SO(4)$ generators commuting with the
$4\times4$ block in \Spmeson\ are ${\ \ \ 1\choose-1\ \ } \otimes{1\ \ 
\choose\ \ 1}$, ${1\ \ \choose\ -1} \otimes {\ \ \ 1\choose-1\ \ }$, and 
${\ \ 1\choose1\ \ } \otimes {\ \ \ 1\choose-1\ \ }$.}
Thus the number of real goldstone modes is ${\cal G}
= n_f(2n_f {-}1) - (n_f{-}2r) (2n_f{-}4r{-}1) - 3r$, and the number of
real parameters describing the Higgs branch is ${\cal P} = r$.  It is a
check on our counting that ${\cal G} {+} {\cal P} = 4{\cal H}$.

It follows from the second $F$-term equation that on an $r$-Higgs
branch $\Phi$ must be zero except in a lower right-hand $(2n_c{-}2r)
\times (2n_c{-}2r)$ block, which can be diagonalized by the unbroken
$\USp(2n_c{-}2r)$ rotations.  A gauge-invariant description of these
$r$-branches is given by a set of constraints on the glue and meson
fields generated by
		\eqn\Spmbconstrii{\eqalign{
0 &= S_\ell\,\, *(M^{n_c-\ell+1}),\qquad 0\le \ell \le n_c ,\cr
0 &= M \cdot M ,\cr
		}}
analogous to \mbconstrii\ and \mbconstriii\ in the $SO(n_c)$ case.

The non-renormalization theorem implies that only the Coulomb factors
of the $r$-branches can be modified quantum mechanically.  We have seen
that classically there exist $r$-branches for $0 \le r \le {\rm
min}\{n_c,n_f/2\}$ which meet the Coulomb branch along submanifolds
with gauge group $\USp(2r)\times U(1)^{n_c-r}$ with $n_f$ fundamental
flavors.  The AF (or finite) microscopic theories have $n_f
\le 2n_c {+} 2$.  Thus, the $\USp(2r)$ factors at the roots of the
$r$-branches are all IR-free (or finite) and so will remain unbroken
quantum-mechanically, with one exception. This is the branch with $r=
[n_f/2]$, which is AF. As we will see, in this case the classical
gauge group is broken quantum mechanically to a maximal subgroup leading
to a non-AF theory. 

Submanifolds of the quantum Coulomb branch with unbroken $\USp(2r)$ gauge
factors are easy to identify explicitly using the exact solution \AS\ in 
terms of Riemann surfaces describing the effective action on the Coulomb 
branch with $n_f>0$ massless flavors:
		\eqn\Spcurv{
y^2 = x\prod_{a=1}^{n_c}(x-\phi^2_a)^2 - 4\Lambda^{2(2n_c+2-n_f)}
x^{n_f-1}.
		}
In the finite case, $n_f = 2n_c {+} 2$, $\Lambda^0$ should be replaced
by a known function of the bare coupling; in the IR-free case,
\Spcurv\ is valid in a sufficiently small neighborhood of $x=\phi_a=0$
where the curve has the simple form $y^2 \propto x^{2n_c+1}$.  Thus,
when $2r{+}1$ branch points of \Spcurv\ coincide, one may expect an
unbroken $\USp(2r)$ gauge symmetry.  On the submanifold with
$r<[n_f/2]$ of the $\phi_a=0$, the curve becomes
		\eqn\Sprootcrv{
y^2 = x^{2r+1}\left[\prod_{a=1}^{n_c-r}(x-\phi_a^2)^2 -
4\Lambda^{2(2n_c+2-n_f)}x^{n_f-2r-2}\right],
		}
giving vacua with an unbroken $\USp(2r)\times U(1)^{n_c-r}$ gauge
group, and so locating the roots of the $r$-branches in the full
quantum theory for $r < [n_f/2]$.\foot{In the special case $r=[n_f/2]$, 
the form of the curve corresponds to the unbroken symmetry
$USp(n_f-2){\times}U(1)^{n_c-n_f/2+1}$ (for $n_f$ even). 
The classical symmetry $USp(n_f)$ is broken quantum mechanically to
$USp(n_f-2)$, which is finite with $n_f$ flavors. In fact, $USp(n_f-2)$
is the maximal subgroup leading to a non-AF theory.}

Though these theories at the
$r$-branch roots are IR--free, quantum effects can change the IR theory
at points where singlets, $e_k$, charged under the $U(1)$ factors become
massless.  Such points are located where the factor in square brackets
in \Sprootcrv\ becomes singular due to pairs of its zeros coinciding.
The maximal such singularity occurs when that factor is a perfect
square, corresponding to $n_c{-}r$ hypermultiplets becoming
simultaneously massless.  As in the $SO(n_c)$ case, there is a single
solution to this condition, namely $r= n_f {-} n_c {-}2$ and $\phi_a^2
= \Lambda^2 (0, \ldots, 0 , \omega, \omega^2, \ldots,
\omega^{2n_c+2-n_f})$ where $\omega = {\rm exp}\{2\pi i/(2n_c{+}
2{-}n_f)\}$.  In an appropriate $U(1)$ basis, the gauge charges of the
light degrees of freedom at this special point are
                \eqn\Spmacro{\matrix{
&\USp(2n_f{-}2n_c{-}4)&\times&U(1)_1&\times&\cdots&\times&U(1)_{2n_c+2-n_f}\cr
2n_f\times Q   &\bf 2n_f{-}2n_c{-}4 && 0      && \cdots && 0      \cr
e_1            &\bf 1             && 1      && \cdots && 0      \cr
\vdots         &\vdots            && \vdots && \ddots && \vdots \cr
e_{2n_c+2-n_f} &\bf 1             && 0      && \cdots && 1      \cr
                }}

We now break to $N{=}1$ supersymmetry by turning on a bare mass for the
adjoint superfield $\Phi$, corresponding to adding a mass term $(\mu/2)
{\rm tr}\Phi^2$ to the superpotential of the microscopic theory.  For
$\mu\gg\Lambda$, we can integrate $\Phi$ out in a weak-coupling approximation,
obtaining an effective superpotential which vanishes as $\mu
\rightarrow\infty$.  At scales above the strong-coupling
scale $\Lambda_1$ of the $N{=}1$ theory, we obtain
$N{=}1$ $\USp(2n_c)$ super--QCD with
$n_f$ flavors\foot{In $N{=}1$ theories we count flavors by {\it pairs}
of squark chiral multiplets since an odd number of fundamental chiral
multiplets is anomalous in $\USp(2n_c)$ \GlobAnom; thus this counting
is the same as the counting of hypermultiplet flavors in the $N{=}2$
theory.} and no superpotential.  If the strong coupling scale
of the $N{=}2$ theory is $\Lambda$, then by a one-loop matching, the
$N{=}1$ scale is $\Lambda_1^{2(3n_c+3-n_f)} = \mu^{2n_c+2}
\Lambda^{2(2n_c+2-n_f)}$.  The appropriate scaling limit sends
$\mu\rightarrow\infty$ and $\Lambda \rightarrow 0$ keeping $\Lambda_1$
fixed, so the model is described by the $N{=}1$ theory at scales
between $\mu$ and $\Lambda_1$, below which the strongly-coupled
dynamics of the $N{=}1$ theory takes over.

We can also study the breaking to $N{=}1$ by beginning with
$\mu\ll\Lambda$.  In this case we should study the effects of $\mu$ on
the $N{=}2$ vacua described above.  It is easy to see that generic
vacua of the $N{=}2$ theory are lifted by nonzero $\mu$; however, the
special point we found on the $r=n_f{-}n_c{-}2\,\ r$-branch is not.
Let $\phi$ denote the adjoint scalar in the $\USp(2r)$ vector
multiplets, and $\psi_k$ the adjoint scalars for each of the $U(1)$
vector multiplets for the unbroken $\USp(2r) \times U(1)^{n_c-r}$
symmetry at the roots of the $r$-branches.  The microscopic mass term
$(\mu/2){\rm tr}\Phi^2$ becomes $\mu(\Lambda\sum_i x_i \psi_i +
{1\over2}{\rm tr}\phi^2 + \ldots)$, where the dots denote higher-order
terms, and $x_i$ are dimensionless numbers.  At any point on an
$r$-branch root for which there are $n_s$ massless singlets, $e_k$,
charged under the $U(1)$'s, with $n_s \le n_c{-}r$, a basis of the
$U(1)$'s can be chosen to diagonalize the charges of the singlets.  The
$F$-terms following from the resulting superpotential have no solution
unless $n_s = n_c{-}r$, showing that only the special vacuum
\Spmacro\ leads to an $N{=}1$ vacuum.  In this case the $e_k$,
$\psi_i$, and $\phi$ fields are massive and can be integrated out,
leaving an effective $N{=}1$ $\USp(2n_f{-}2n_c{-}4)$ super--QCD with
$n_f$ flavors.  This is precisely the dual gauge group of \Sei\IP.

\vfill\eject
\centerline{{\bf Acknowledgments}}

It is a pleasure to thank M. Alford, L. Dixon, N. Seiberg, and
M. Strassler for helpful discussions and comments. This work was supported
in part by DOE grants DE-FG05-90ER40559 and DE-FG02-91ER75661, the
Center for Basic Interactions, and an Alfred P. Sloan fellowship.
M.R.P. thanks the Rutgers physics department for hospitality while
part of this work was completed.

\listrefs
\end